\documentclass[prd,twocolumn,letterpaper,superscriptaddress]{revtex4}
\include{epsf}
\usepackage{graphicx}
\usepackage{pslatex}
\usepackage{url}

\begin{document}

\title{Erratum: Observational Constraints on the Ultra-high Energy Cosmic Neutrino Flux\\
from the Second Flight of the ANITA Experiment}


\author{P.~W.~Gorham$^1$,
P.~Allison$^1$,
B.~M.~Baughman$^2$,
J.~J.~Beatty$^2$,
K.~Belov$^3$, 
D.~Z.~Besson$^4$,
S.~Bevan$^5$,
W.~R.~Binns$^6$,
C.~Chen$^7$,
P.~Chen$^7$,
J.~M.~Clem$^8$,
A.~Connolly$^5$,
M.~Detrixhe$^4$,
D.~De~Marco$^8$,
P.~F.~Dowkontt$^6$,
M.~DuVernois$^1$,
E.~W.~Grashorn$^2$,
B.~Hill$^1$,
S.~Hoover$^3$,
M.~Huang$^7$
M.~H.~Israel$^6$,
A.~Javaid$^8$,
K.~M.~Liewer$^{9}$,
S.~Matsuno$^{1}$,
B.~C.~Mercurio$^2$,
C.~Miki$^{1}$,
M.~Mottram$^5$,
J.~Nam$^7$,
R.~J.~Nichol$^5$,
K.~Palladino$^2$,
A.~Romero-Wolf$^1$,
L.~Ruckman$^1$,
D.~Saltzberg$^3$,
D.~Seckel$^8$,
R.Y.~Shang$^7$,
G.~S.~Varner$^{1}$,
A.~G.~Vieregg$^3$,
Y.~Wang$^7$
}
\vspace{2mm}
\noindent
\affiliation{
$^1$Dept. of Physics and Astronomy, Univ. of Hawaii, Manoa, HI 96822. 
$^2$Dept. of Physics, Ohio State Univ., Columbus, OH 43210. 
$^3$Dept. of Physics and Astronomy, Univ. of California Los Angeles, CA 90095.
$^4$Dept. of Physics and Astronomy, Univ. of Kansas, Lawrence, KS 66045. 
$^5$Dept. of Physics and Astronomy, Univ. College London, London, United Kingdom.
$^6$Dept. of Physics, Washington Univ. in St. Louis, MO 63130. 
$^7$Dept. of Physics, 
National Taiwan Univ., Taipei, Taiwan.
$^8$Dept. of Physics, Univ. of Delaware, Newark, DE 19716. 
$^9$Jet Propulsion Laboratory, Pasadena, CA 91109. 
}

\maketitle

In a recent article~\cite{Anita10} we reported a limit on the
cosmic neutrino flux from the second flight of the ANITA
experiment.  The limit was based on observing two events passing all
cuts on a background of $0.97\pm 0.42$.

One of the first steps 
in the blind analysis procedure was inserting twelve pulser
events at undisclosed
random times to mimic a neutrino signal. These events would be removed 
upon unblinding the analysis. This was one of two ways
that the analysis employed a blind analysis technique.
After publication, we subsequently determined that due to a clerical error
one of the two surviving
events, Event 8381355, was actually one of the inserted pulser events.  
The fact that this event survived its subsequent scrutiny we
consider as a demonstration that the blinding procedure was truly valid.

\begin{figure}[ht]
\begin{center}
\includegraphics[width=3.2in]{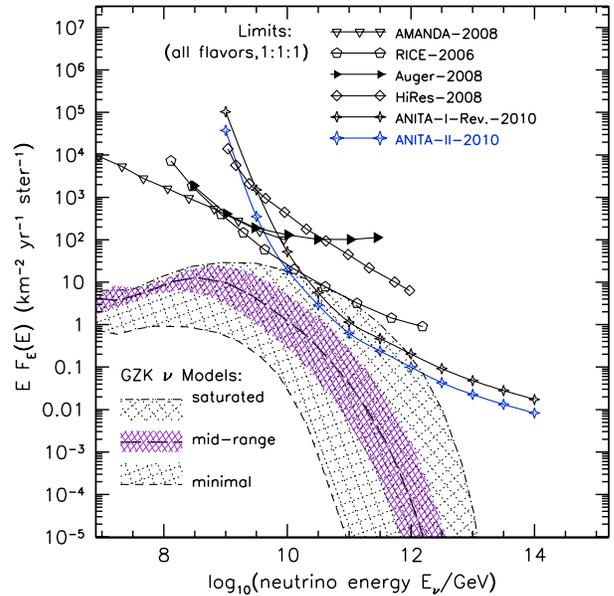} \\
\caption{ANITA-II limit for 28.5 days livetime. The blue
curve is the new actual limit, based on the one surviving candidate. Other
limits are from AMANDA, RICE,
Auger, HiRes, and a revised limit from ANITA-I. 
The BZ (GZK) neutrino model range is determined by a variety of 
models. Full citations are given in the original article.}
\label{lim10}
\end{center}
\end{figure}

The net result is that ANITA-II observed one event on a background 
of $0.97\pm0.42$.  
The new limit, which is 33-34\% stronger, is 
shown in in Figure~\ref{lim10}.
Now the actual limit is essentially the same as the expected limit so
we no longer show both curves.
The ANITA-II 90\% CL integral flux limit on a pure $E^{-2}$ spectrum
for $10^{18}$~eV $\leq E_{\nu} \leq 10^{23.5}$~eV is 
$E_{\nu}^2 F_{\nu} \leq 1.3 \times 10^{-7}$~GeV~cm$^{-2}$~s$^{-1}$~sr$^{-1}$. 
An updated evaluation of confidence limits for constraining representative models is given in
Table~\ref{model-table}.  
The changes result in an improvement in the 
constraints on the given strong-source 
evolutionary models, the majority of which are now excluded at
$>90$\% confidence.

\begin{table}[hbt!]
 \begin{footnotesize}
  \begin{tabular}{lcr}
\hline
{ {\bf Model \& references}}   &  predicted {$N_{\nu}$}~~      &  ~~{\bf CL,\%} \\ \hline
{\it Baseline models:} &  &  \\
~~~~~~Various & 0.3-1.0 & ...\\
{\it Strong source evolution models:}&  &  \\
~~~~~~Aramo {\it et al.} 2005 & 2.4 & 85 \\
~~~~~~Berezinsky 2005 & 5.1 &   98 \\
~~~~~~Kalashev {\it et al.} 2002 &  5.6 & 99 \\
~~~~~~Barger, Huber, \& Marfatia 2006 & 3.5 & 93 \\
~~~~~~Yuksel \& Kistler 2007 & 1.7 & 74 \\
{\it Models that saturate all bounds}: & & \\
~~~~~~Yoshida {\it et al.} 1997 & 30 & $>99.999$  \\
~~~~~~Kalashev {\it et al.} 2002 &  19 & $>99.999$ \\
~~~~~~Aramo {\it et al.} 2005 & 16 & 99.999 \\
{\it Waxman-Bahcall fluxes}: & & \\
~~~~~~Waxman, Bahcall 1999, evolved sources~~ & 1.4 & ...  \\
~~~~~~Waxman, Bahcall 1999, standard & 0.5 & ...   \\ \hline
  \end{tabular}
\caption{Expected numbers of events $N_{\nu}$ from several cosmogenic neutrino models, and 
confidence levels for exclusion by ANITA-II observations when
appropriate.  Citations are given in the original article.
\label{model-table}}
 \end{footnotesize}
\end{table}

\end{document}